\documentclass[useAMS,usenatbib,usegraphicx]{mn2e}
\usepackage{array}

\newcommand{\spitzer}{{\it Spitzer}}
\newcommand{\galex}{{\it GALEX}}
\newcommand{\hst}{{\it HST}}
\newcommand{\mass}{{\it 2MASS}}
\newcommand{\mi}{\,$\mu$m}
\newcommand{\av}{{$A_{\rm V}$}}
\newcommand{\ha}{H$\alpha$}

\newcommand{\msun}{\,M$_\odot$}

\title[Star Formation Histories]{Star Formation Histories within the Antennae Galaxies (Arp~244)}
\author[Hong-Xin Zhang, Yu Gao and Xu Kong]
  {Hong-Xin Zhang$^{1,2,4}$, Yu Gao$^{1}$\thanks{E-mail: yugao@pmo.ac.cn} 
   and Xu Kong$^{3}$\\
   $^{1}$Purple Mountain Observatory, Chinese Academy of Sciences, 2
   West Beijing Road, Nanjing, Jiangsu 210008, China\\
   $^{2}$Graduate School of the Chinese Academy of Sciences, Beijing 100080, China\\
   $^{3}$Center for Astrophysics, University of Science and Technology of China, Hefei 230026, China\\
   $^{4}$Current affiliation: Lowell Observatory, 1400 West Mars Hill Road, Flagstaff, AZ 86001, USA}
\begin{document}
\date{Accepted 2009 September 21.  Received 2009 September 7; in original form 2009 May 14}
\pagerange{\pageref{firstpage}--\pageref{lastpage}} \pubyear{2009}
\maketitle
\label{firstpage}
\begin{abstract}
With the imagery from \galex, \hst, \mass, and \spitzer, 
and at the resolution of MIPS~24 \mi~($\sim$ 6$''$), we study the  variations 
of the broadband spectral energy distributions (SEDs) of star-forming regions
within the nearest prototypal major merger --- the Antennae galaxies. 
By including MIPS~24\mi~dust emission into stellar population analysis, 
we reliably, albeit roughly, constrain the star formation histories of these 
24\mi~selected star-forming regions across the merging disks of the Antennae. 
Our population analysis is consistent with the star formation scenario that, 
most regions across the whole system are at a modest level of star formation with 
the exception of some localized intense starburst sites in the well-known {\it overlap} 
regions and the {\it western-loop} regions of northern galaxy NGC~4038. 
Compared with all the other regions, the young {\it overlap} regions currently ($<$ 10 Myr) 
are experiencing much more violent enhancement of star formation. 
Across the {\it overlap} regions, we suggest two sequential star formation 
paths which we interpret as the imprints of the interpenetrating process of the two merging 
disks following their second close encounter. And we suggest that the star formation in the 
southern and (especially) northwestern edges of the {\it overlap} zone may have been just 
triggered by pre-starburst shocks. 
The well-known mid-infrared ``{\it hotspot}'' in the {\it overlap} regions is also 
a ``{\it hotspot}'' at 4.5\mi, whose total 4.5\mi~emission ($\ge$ 80\% from both 
hot dust and atomic/molecular lines) is comparable with that of the two galactic nuclei. 

\end{abstract}

\begin{keywords}
galaxies: individual (the ``Antennae'' galaxies) -- galaxies: interactions -- galaxies: stellar content -- galaxies: starburst -- galaxies: photometry 
\end{keywords}

\section{Introduction} 
Galaxy mergers, expecially major mergers, can dramatically influence the morphological 
and star-forming properties of galaxies over relatively short timescales. 
Almost all the ultraluminous infrared galaxies (ULIRGs)
--- the strongest starbursts in the local Universe, are in interacting/merging 
systems \citep{Sanders1996}. Moreover, \citet{conselice2003b} suggested that about two 
thirds of submillimeter galaxies at z $>$ 1 are undergoing major mergers. 
Galaxy interactions/mergers seem to be very frequent in the past 
(e.g. \citealt{lefevre2000}; \citealt{patton2002}; \citealt{conselice2003a}; 
\citealt{Elbaz2003}; \citealt{kartaltepe2007}; \citealt{deravel2008}; 
\citealt{lin2008}; \citealt{conselice2009}). 
Therefore, it is of great importance to understand how the burst of star formation is 
triggered in the course of interacting/merging. 

At a distance of 19.2 Mpc (H$_{0}$ = 75 km s$^{-1}$ Mpc$^{-1}$)
\footnote{We note the recent debate about the distance to the Antennae. 
\citet{saviane2008} determined a distance of $\sim$ 13.3 Mpc from the tip 
of red giant branch, whereas \citet{schweizer2008} estimated a distance of 
$\sim$ 22.3 Mpc based on the type Ia supernovae 2007sr light curve. 
Throughout this work, we assumed the traditionally adopted Hubble Flow distance. 
However, our conclusions are not affected by the controversy over the distance.}
, the Antennae (NGC~4038/39, Arp~244) is the nearest prototypal major merger
between two gas-rich spiral galaxies (\citealt{Toomre1972}; \citealt{Hibbard2001}).
Thus it provides us with a unique opportunity to study the induced star formation
process as a consquence of interaction in detail. It has been extensively studied at
essentially all wavelengths from X-ray to radio (
\citealt{Hummel1986}; \citealt{Read1995}; \citealt{Vigroux1996}; \citealt{Mirabel1998}; 
\citealt{Nikola1998}; \citealt{Whitmore1999}; \citealt{Neff2000}; \citealt{Wilson2000}; 
\citealt{fabbiano2001}; \citealt{Gao2001}; \citealt{Hibbard2001}; \citealt{fabbiano2003}; 
\citealt{fabbiano2004}; \citealt{Wang2004}; \citealt{Hibbard2005}; \citealt{Mengel2005}; 
\citealt{bastian2006}; \citealt{Gilbert2007}; \citealt{Schulz2007}; \citealt{brandl2009}). 
{\it ISO} mid-infrared (MIR) observations \citep{Mirabel1998} show that the most intense starburst
in this system takes place in the so-called {\it overlap} region between the 
two nuclei, which indicates its intermediate merging stage, almost totally obscured 
in the optical. 
Considering the abundant molecular gas, wide spread star formation and
overall modest star formation efficiency, Gao~et al. (2001) argued that
Arp~244 has the potential of producing an ultraluminous extreme starburst 
in a later stage of merging.

In the Antennae, observations with \hst~have identified thousands of super star clusters (SSCs)
\citep{Whitmore1999} possibly being formed as part of the merging process.
Both theoretical predictions \citep{Goodwin2006} and observations
(\citealt{Whitmore2004}; \citealt{fall2005}; \citealt{Mengel2005}) suggest that many of these SSCs 
will dissolve rapidly into the field stellar population
in the course of galaxy-galaxy merging. This process was coined as star cluster's
``{\it infant mortality}'' \citep{Whitmore2004}. Furthermore, a large number of stars 
may not form in star cluster mode. In fact, star clusters found in the deep \hst~images  
only contribute 9\%, 8\%, 5\%, and 7\% of the apparent total {\it U, B, V,} and {\it I} band 
light of the Antennae \citep{Whitmore2002}. 
Due to the uncertainties on the cluster formation history and efficiency, recently, 
\citet{bastian2009} pointed out the difficulty in the accurate understanding of 
the age distribution of star clusters in mergers like the Antennae. 
Therefore, in order to obtain a complete picture of star formation histories in 
the course of merging, we must study the extended stellar populations, in addition 
to these star clusters.

Evolutionary population synthesis has become a powerful tool of  
interpretation of the integrated spectrophotometric observations of 
galaxies and sub-galactic regions. 
The most common method of model-observation comparison 
for stellar population analysis in galaxies 
is SED-fitting, 
with either least-squares or chi-squared minimization technique  
(e.g. \citealt{kong2000}; \citealt{Gavazzi2002}). However, the well-known 
age-extinction degeneracy problem prevents us from obtaining 
reliable information about the star formation history for these galaxies or 
their sub-galactic regions, especially when only broadband photometry data are available.
This is because these extended regions have an unknown mixture of various stellar 
populations, and the different populations may experience totally different extinctions 
(\citealt{calzetti1994}; \citealt{Charlot2000}).

With the inclusion of the high-quality and high resolution \spitzer\ 24\mi~dust emission data 
in our population analysis, we show here that the
degeneracy between stellar population and dust extinction can be broken to a great extent. 
Thus, for the first time, 
we can reliably, albeit roughly, constrain the star formation histories within the Antennae galaxies
using SEDs over the whole spectral range from far-UV (FUV) to MIR. 
The outline of this paper is the following: In
Sect.~\ref{sec:data} we introduce the multi-band data that we use in this study, and give the 
multiwavelength photometry of star-forming regions selected mainly from the 24\mi~image. 
Sect.~\ref{sec:sed} gives some brief comparisons of the broadband SEDs and their
variations across the whole system. 
Sect.~\ref{sec:mod} presents our methodology to constrain the star formation histories 
across the merging disks, and the main results of our population analysis.
We discuss these results in Sect.~\ref{sec:dis} and then a
summary of our main findings follows in Sect.~\ref{sec:sumr}.

\section{Data and Photometry} \label{sec:data}

\subsection{Data}
Both FUV ($\sim$ 1516~\AA) and near-UV (NUV; $\sim$ 2267~\AA) images were derived 
from the \galex~~Ultraviolet Atlas of Nearby Galaxies distributed by \citet{GildePaz2007}. 
The FWHMs of the PSFs are 5$''$ and 6$''$ at FUV and NUV, respectively. With these data
\citet{Hibbard2005} have studied the stellar populations of the famous tidal regions.

Four broadband (F336W, F439W, F555W, F814W) and one narrowband (F658N; \ha)
images \citep{Whitmore1999} taken with the WFPC2 aboard
the \hst~were obtained as B associations from the MAST Archive 
\footnote{
http://archive.stsci.edu/hst/wfpc2/search.html
}, and mosaic containing the four chips for each image was created.

NIR {\it JHKs} atlas images from {\it 2MASS}~were retrieved through the
Interactive {\it 2MASS}~Image Service
\footnote{
http://irsa.ipac.caltech.edu/applications/2MASS/IM/interactive.html
}.

The \spitzer~imagery (3.6, 4.5, 5.8, 8.0, 24\mi) of the Antennae 
was obtained with both the Infrared Array Camera (IRAC; \citealt{Fazio2004}) 
and the Multiband Imaging Photometer for \spitzer~(MIPS; \citealt{Rieke2004}) 
on board the {\it Spitzer Space Telescope}. The Basic Calibrated Data (BCDs) were 
retrieved with the Leopard software 
\footnote{
Available at http://ssc.spitzer.caltech.edu/propkit/spot/
}
. Background matching, cosmic-ray removal,
flat-fielding and mosaicking were performed using the Spitzer Science Center's reduction 
software package MOPEX 
\footnote{
http://ssc.spitzer.caltech.edu/postbcd/download-mopex.html
}.
Images of the four IRAC bands have previously been presented by \citet{Wang2004}.

\subsection{Region Selection and Photometry Extraction}

Prior to our multiwavelength photometry comparison and extraction, all images 
were background removed, registered/aligned, and resampled to the same pixel scale (1.5$''$).
Then all images (except FUV/NUV) are convolved to the same resolution of MIPS~24\mi~with 
the convolution kernels provided by \citet{Gordon2008}.

\begin{figure*}
\centering
\includegraphics[width=0.495\textwidth]{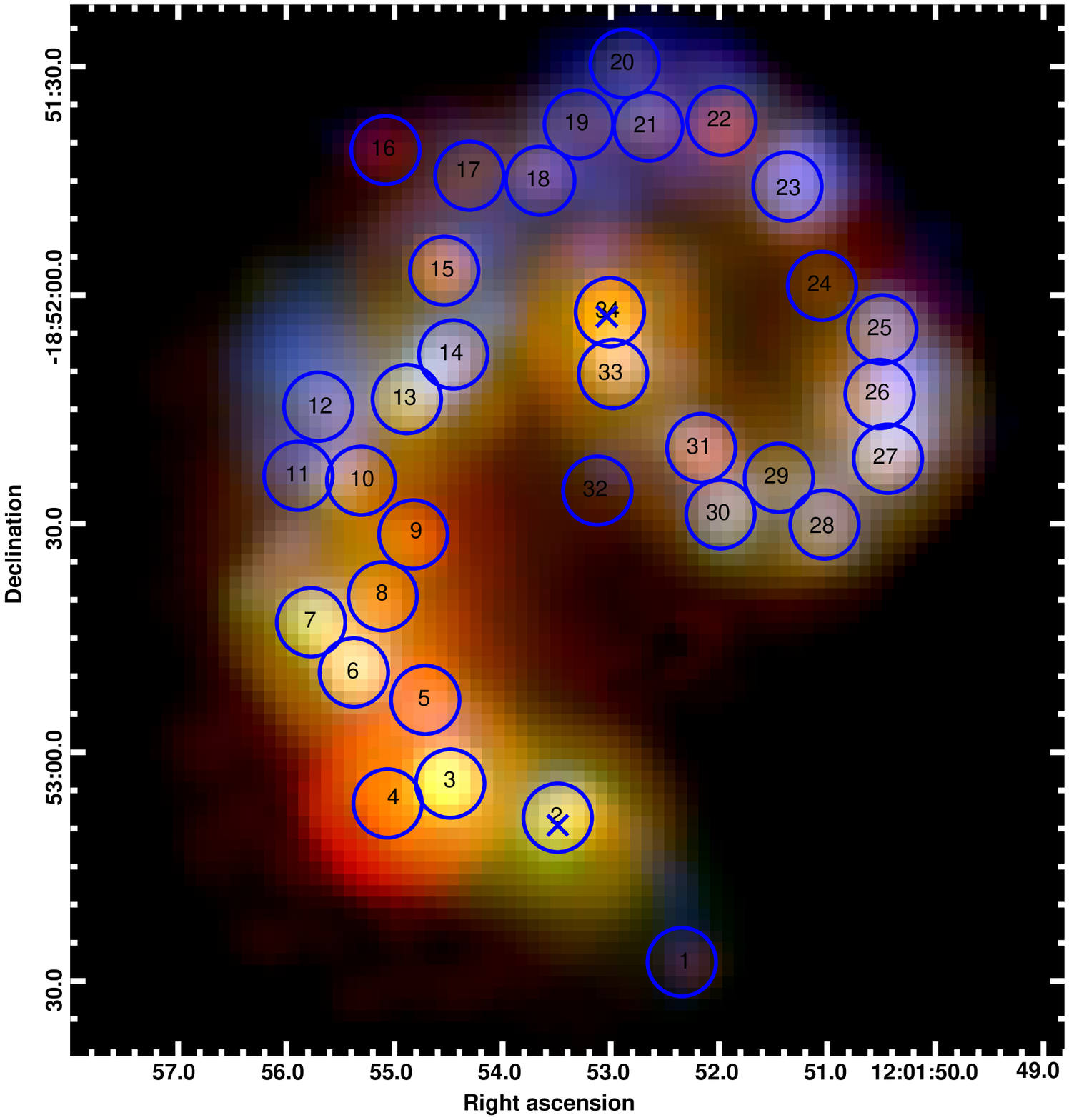}
\includegraphics[width=0.495\textwidth]{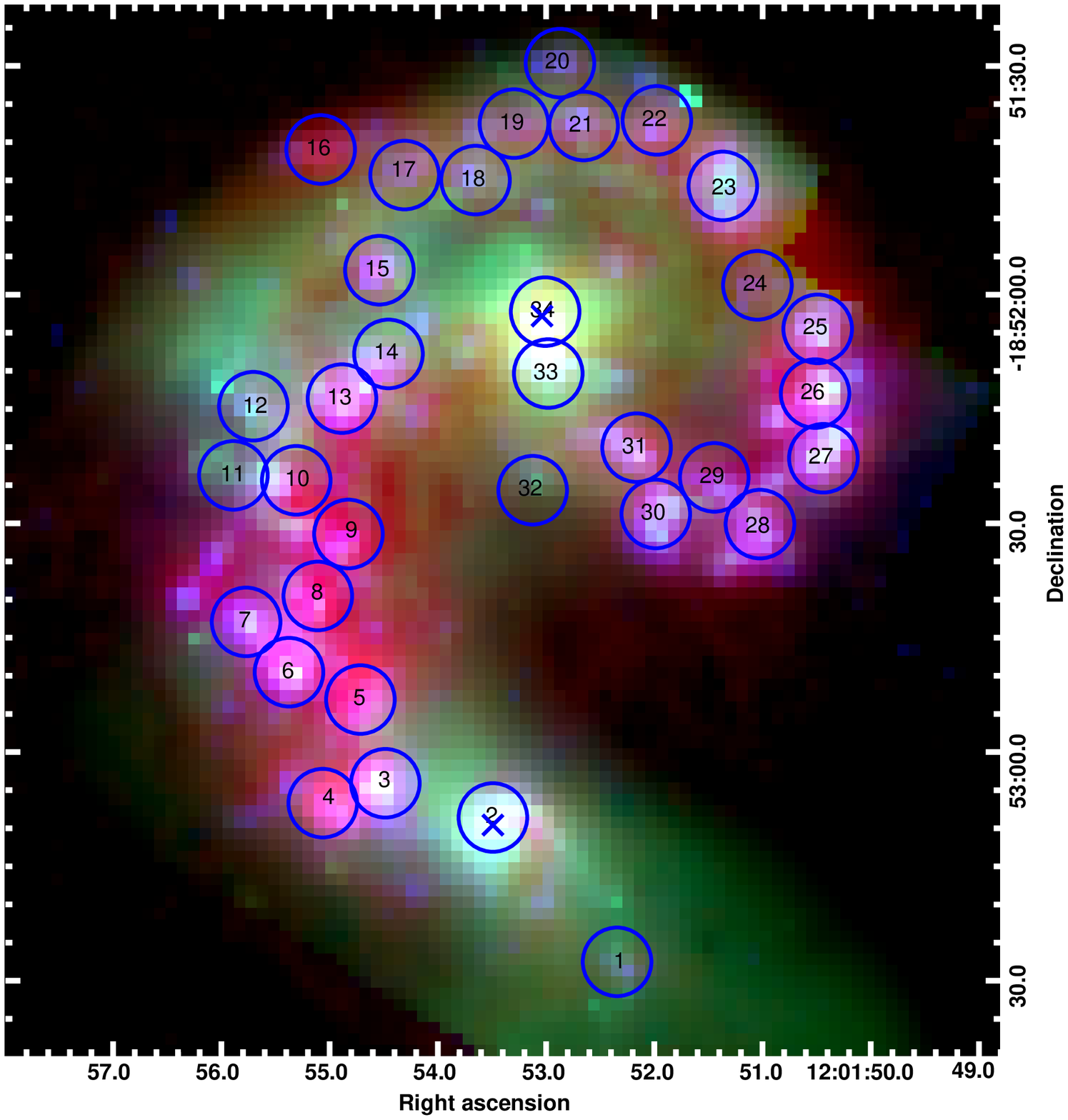}
\caption{
{\itshape Left:\/} 
Circular apertures of 9$''$ in diameter were selected mainly as 24\mi~peaks and 
superposed on color-composite images generated from MIPS~24\mi~(\textit{red}), 
\hst~\ha~(\textit{green}), and \galex~FUV(\textit{blue}) maps of the Antennae.  
The \hst~\ha~map has been degraded to match the resolution of MIPS~24\mi. 
The {\it cross} signs mark the nuclear centers of the two colliding galaxies.
Pixel scales are 1.5$''$. {\itshape Right:\/} Color composite image generated from the
IRAC~8\mi~(\textit{red}), \hst~F814W~(\textit{green}) and \ha~(\textit{blue}) maps
at their original resolutions (re-sampled to 1.5$''$ pixel scales). 
The pixel scales of the two figures are 1.5$''$. 
The original resolution of IRAC 8\mi~is $\sim 2''$. The mosaicked \hst~images 
originally have $\sim$ 0.1$''$ pixels (undersampled). 
At the adopted distance, 1$''$ corresponds to $\sim$ 93 pc.} 
\label{fig:img}
\end{figure*}

We select star-forming regions primarily as 24\mi~emission peaks, since 
24\mi~has been shown to be a very good local tracer of current star formation 
(e.g. \citealt{calzetti2005}). 
Practically, we first use the {\it IRAF} 
\footnote{
{\it IRAF} is distributed by the National Optical Astronomy Observatories 
which are operated by the Association of Universities for Research
in Astronomy, Inc., under cooperative agreement with the National
Science Foundation
}
DAOFIND task in DAOPHOT stellar photometry package \citep{Stetson1987}
to find all local emission enhancements with 10 $\sigma$ signal-to-noise (S/N) ratio threshold 
in the 24\mi~image. Then we examine the findings of local enhancements visually and 
check carefully to ensure that these previously 
detected enhancements are true star-forming regions rather than artifacts due to the Airy 
diffraction rings or spikes. We also include three additional FUV bright regions 
without any nearby 24\mi~peaks associated (regions 11,12 and 32). 
Finally, a total of 34 nearly non-overlapping circular apertures  of 9$''$ ($\sim$ 800~pc) 
in diameter (Figure~\ref{fig:img}) were selected. The large aperture size is mainly dictated 
by the PSF of 24\mi~and the rather extended emission features for most regions. 

Our aperture size for photometry is, in any sense, significantly larger than a typical 
H$_{~\rm II}$ region (\citealt{Knapen1998}; \citealt{Oey2003}), and even larger than a typical central star-forming region of a starburst galaxy. 
However, as is already found by \citet{zhang2001}, the spatial distribution of the young star 
clusters tends to be correlated up to physical scale of $\sim$ Kpc.  
\citet{bastian2006} have studied several star cluster complexes with sizes up to several hundred 
parsecs in the Antennae, they found that the young cluster complexes often share the same general 
velocity distribution with associated giant molecular clouds (GMCs), and even some complexes 
themselves are clustered. Our star-forming regions are usually spatially resolved into one 
or few such bright star cluster complexes dominated by few bright star clusters on the 
high-resolution optical/NIR imaging (e.g. \citealt{Whitmore1999}; \citealt{Mengel2005}).  
Therefore, most of our regions should be supposed to be complexes of star clusters over 
extended background stellar populations. 
We note that some regions show slight, yet not systematic, displacement ($\sim1-2''$) between the 
24\mi~and the associated FUV peaks. The displacement may indicate that IR and UV emission are 
dominated by different clusters within these regions. 

Although the measured FWHMs of the \galex~PSFs are close to that 
of the PSFs in MIPS 24\mi~image, we apply point source aperture corrections to
FUV (1.206), NUV (1.269) and other bands~(1.982). 
Aperture corrections are determined from photometry with increasing 
aperture radii of isolated field stars in the image fields. 
All the fluxes have been corrected for foreground Galactic extinction using
a value of {\it E(B-V)} = 0.046 \citep{Schlegel1998} and \citet{cardelli1989}
extinction law with R$_{V}$=3.1. 
The uncertainties assigned to the photometric values are a quadratic
sum of three contributions: variance of the local background, photometric
calibration uncertainty (5\% for FUV and WFPC2, 3\% for NUV,
10\% for IRAC, and 4\% for 24\mi), and poisson statistics noise.

\section{Broadband SEDs} \label{sec:sed}

\begin{figure*}
\includegraphics[width=175mm]{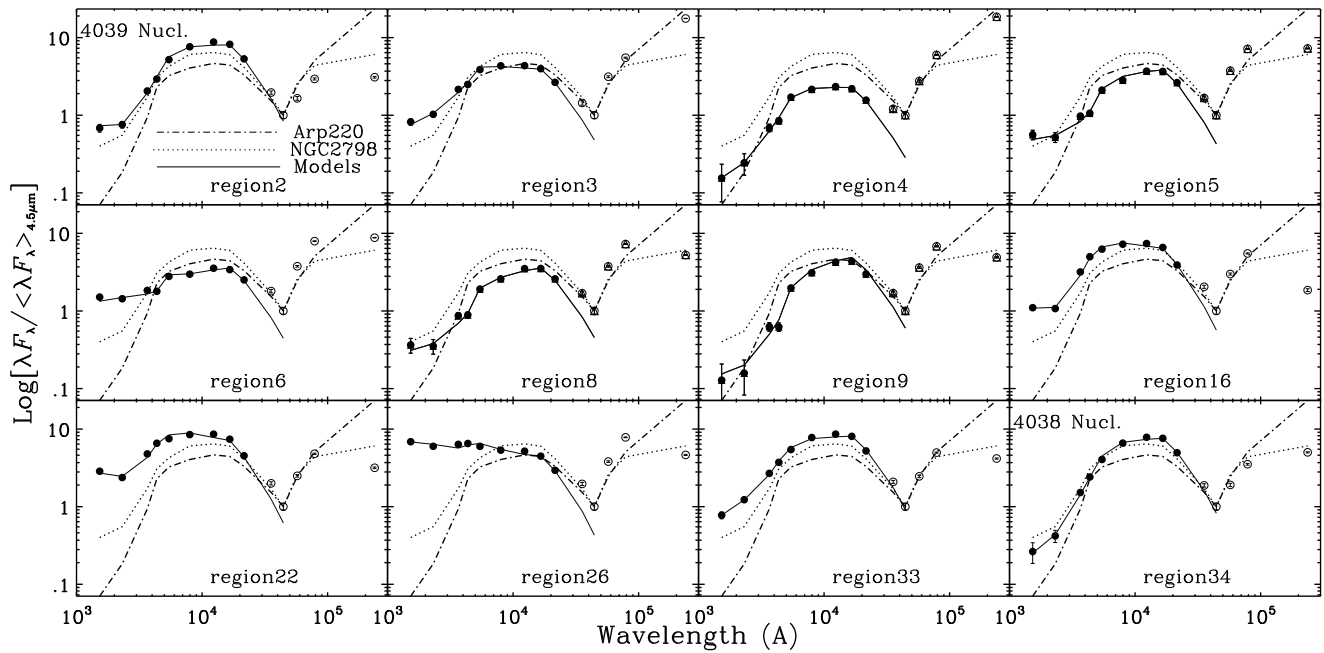}
\caption{Display of FUV-24\mi~broadband SEDs of some representative star-forming regions 
marked in Figure~\ref{fig:img} and compared to Arp~220 and local H$_{~\rm II}$  
galaxy NGC~2798. The broadband photometry data points from FUV to {\it Ks}, which are fitted 
with 3-SSP superpositions, are represented by \textit{filled circles} and 
the \spitzer~data points by \textit{open circles}.
For the purpose of comparison, the broadband WFPC2 data have been transformed to 
standard {\it UBVI} system with the transformation coefficients provided by \citet{holtzman1995}. 
The best-fit (FUV-{\it Ks}) composite stellar broadband SED models have been plotted 
as \textit{solid lines}.} 
\label{fig:sed1}
\end{figure*}

Figure~\ref{fig:sed1} shows the broadband SEDs for some representative regions across 
the Antennae galaxies as compared to that of entire Arp~220 and the local H$_{~\rm II}$    
galaxy NGC~2798 \citep{Kinney1993}. There are some 
very remarkable differences of the SEDs between different regions across the whole system
of the Antennae galaxies. 

In the IR dust emission bands, 
24\mi~very small grains (VSGs) and 8\mi~PAH emission (for high-metallicity regions) all show 
very good correlation with extinction-corrected hydrogen recombination 
emission from starburst regions of normal galaxies to luminous IR galaxies (LIRGs) and
ULIRGs (\citealt{alonso2006}; \citealt{calzetti2007}). Nevertheless, 
the 8\mi~PAH emission is progressively depressed relative to 24\mi~with increasing 
star formation intensities (e.g. \citealt{calzetti2005}; \citealt{Smith2006}; \citealt{Draine2007}; 
\citealt{Thilker2007b}). 
Hence, the 24\mi/8\mi~ratio would give a rough manifestation of different strengths of 
{\it current} star formation activities across the whole system. 

Obviously, the {\it overlap} regions (3 - 9) generally have higher 24\mi/8\mi~ratios than  
other regions in the Antennae, meaning that the {\it overlap} regions are the most 
intense {\it current} star-forming sites. \spitzer~IRS spectrum \citep{brandl2009} 
also show that the {\it overlap} regions have overall the weakest relative strength (normalized 
to 15\mi~continuum flux) of PAH features among all these star-forming regions. 
High-resolution Optical and NIR observations (\citealt{Whitmore1999}; \citealt{Mengel2005}) 
suggest that the {\it overlap} regions host most of the youngest clusters. 
Region 4, which hosts the well-known 
MIR ``{\it hotspot}'' \citep{Mirabel1998} of the {\it overlap} regions, 
has the highest 24\mi/8\mi~ratios, even comparable with that of Arp~220. 
High-quality MIR spectrum \citep{brandl2009} obtained from \spitzer~IRS more clearly 
show that region 4 is characterized by very hot dust emission and is among the regions with the 
strongest strength of the radiation field. 
The intrinsically brightest star cluster (WS95-80, \citealt{whitmore1995}) coincides 
with region 4.

The regions (16 - 22) in the {\it northern} galaxy NGC~4038 have the lowest 24\mi/8\mi~ratios 
and reddest {\it UB} colors as compared with other regions, which is consistent with the fact 
that large number of rather old (100 Myr $\sim$ 500 Myr) star clusters are found there 
(\citealt{Whitmore1999}; \citealt{Mengel2005})

\begin{figure*}
\includegraphics[width=175mm]{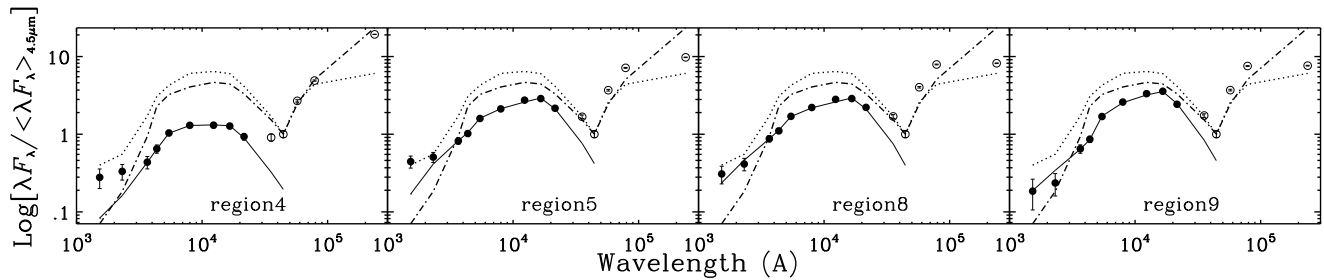}
\caption{Display of FUV-24\mi~broadband SEDs (extracted from the images 
with their {\it original resolutions}) of four {\it overlap} star-forming regions, 
which may have significant starlight contamination by nearby bright regions at short 
wavebands. Lines and symbols are the same as in Figure~\ref{fig:sed1}, except that FUV/NUV data 
are not used in the SED-fittings. 
It can be seen that the overall shapes of the SEDs do not change as compared with the SEDs 
extracted from the smoothed images (Figure~\ref{fig:sed1}).} 
\label{fig:sed2}
\end{figure*}

In the UV wavebands, on the contrary, the {\it Western-loop} regions
(22-32) have the strongest UV emissions, largest UV excesses, 
and relatively strong 8\mi~(PAH) emissions, yet low 24\mi/8\mi~ratios. 
These are in contrast with those in the overlap regions and indicate their 
relatively late star formation stage, which is also evidenced by the large \ha~bubbles 
found there \citep{Whitmore1999}. \citet{brandl2009} also presents the IRS spectra of the 
UV brightest region 26, which has significantly weaker strength of 
the radiation field, yet stronger PAHs emission features, than do the {\it overlap} regions, 
in agreement with our broadband analysis.  

The circumnuclear star-forming region of NGC~4038, namely region 33, has 
24\mi/8\mi~ratio close to that of the {\it overlap} regions, suggesting its intense 
{\it current} star formation activity. The prominent, yet redder, UV emission reveals that
this region may also hold large post-starburst ($>$ 10~Myr) stellar populations. 
For the two nuclei regions, NGC~4039 has both weaker 8\mi~PAH emission 
and 24\mi~dust emission indicating its lower star formation activity, 
whereas NGC~4038 has almost the same broadband SEDs as that of the local H$_{~\rm II}$ galaxy NGC~2798 in the plotted wavelength range. 
Thus far there is no strong evidence for the presence of AGN activity in the two nuclei. 
The IRS spectrum (\citealt{brandl2009}) suggest that, 
both nuclei regions have significantly weaker strength of the radiation field than the 
{\it overlap} regions. 
In comparison with the southern nucleus region, 
the northern nucleus region has flatter MIR continuum, stronger PAHs fluxes, yet smaller 
PAHs equivalent widths. These are consistent with our broadband analysis, i.e.   
both nuclei regions have weak {\it current} star formation strength 
compared to the {\it overlap} starburst regions, 
and the southern nucleus region has even weaker current star formation strength than does 
the northern nucleus. 
The stronger high excitation lines in the southern nucleus region \citep{brandl2009} should 
be attributed to supernovae shocks, which are also strongly evidenced by its very steep radio 
spectrum \citep{Neff2000}. 
We point out that our apertures covering the two nuclei (regions 2, 34) are so large that
the surrounding star-forming regions are also included.

When we examine the SEDs variations across the {\it overlap} regions, we see an interesting 
phenomenon. From the central regions, i.e. regions 6 and 7, to both the northwestern (e.g. regions 8, 9) 
and southeastern (e.g. regions 4) regions, the UV/optical emission gradually become redder and 
weaker, accompanying with comparable amount of energy fraction re-emitted as IR dust emission. 
In the following sections, we show that the obvious trends of the SEDs variations 
across the {\it overlap} regions may reflect the star formation modes in the {\it overlap} regions.

\section{Stellar Population Analysis} \label{sec:mod}

\subsection{Methodology} 
The availability of \ha~and MIPS~24\mi, which traces the star formation (unobscured and obscured) on 
timescales of $\sim$ 10~Myr, FUV and NUV (tracing star formation on timescales of $\sim$ 100~Myr), 
and the {\it U}(F336W) and {\it B}(F439W), whose combination in {\it U}/{\it B} ratio is sensitive 
(straddles the 4000\AA~break) to the fraction of populations younger than a few hundred Myr relative to 
the old populations, makes it reasonable to roughly probe the star formation histories across 
the whole merging disks of the Antennae. 
With $\chi^{2}$ minimization technique, we fit the broadband SEDs (FUV-{\it Ks}) plus 
the narrowband \ha~photometry for the above selected 
star-forming regions with superpostions of three single stellar populations 
(SSPs), namely young ($<$ 10~Myr), intermediate (10~Myr-300~Myr) and 
old ($>$ 300~Myr) populations. In our fitting, we account for the different extinctions 
experienced by different populations. 

For population synthesis models, 
we use the new version Starburst99 \citep{Vazquez2005} SSP models 
with the Padova~2000 stellar evolution tracks, which include the full AGB 
evolution. 
We adopt the multi-power law Kroupa initial mass function \citep{Kroupa2002}. 
Adoption of the bottom-heavy Salpeter IMF would not change our main conclusions 
in this paper. Using the metallicity sensitive Mg~I line at 8806.8 \AA,
\citet{Mengel2002} found a handful of star clusters in Arp~244
with solar metallicity, which is consistent with recent metallicity estimation 
(\citealt{bastian2009}) from more absorption and emission lines in the Antennae. 
Thus, we fix solar metallicity for all the SSPs models.
Since we have narrow-band \ha~imagery data, and the regions we studied
are all actively star-forming regions, we include gaseous emission into the SSP
models. Besides the nebular continuum emission and several hydrogen emission lines
provided by Starburst99, we also account for other strong hydrogen emission lines 
(i.e. Balmer, Paschen, Brackett and Pfund lines) assuming case B recombination and strong 
non-hydrogen element emission lines using the line ratios of typical Galactic 
H$_{~\rm II}$ regions compiled by \citet{Anders2003}. 

For dust extinction,
we use the two-phase dust attenuation recipes developed by \citet{Charlot2000}. 
The effective absorption curve is proportional to $\lambda^{-0.7}$. 
One adjustable parameter $\mu$ defines the ratio of the total effective
extinction experienced by intermediate/old ($>$ 10~Myr) to young ($<$ 10~Myr)
populations. 

It is becoming well known, from the works of the SINGS team, that the combination of \ha\ 
and 24\mi\ accounts for both the unobscured and obscured current star formation for 
star-forming regions in galaxies (\citealt{calzetti2007}; \citealt{kennicutt2007}). 
We adopt the relationship between extinction-corrected \ha,~observed 
\ha~and 24\mi~calibrated for the H$_{~\rm II}$ regions in the 33 SINGS galaxies  
(equation 5 in \citet{calzetti2007}) to obtain the \ha\ extinction, and then use the 
\citet{Charlot2000} extinction curve to derive the approximate \av~of the young stellar 
populations from the following equation
\begin{equation}
A_{\rm V} = 2.825\log\left[1+0.031\frac{L_{24\mu m}}{L_{H\alpha}}\right]
\end{equation} \label{eqn:eq1}
where $L_{24\mu m}$ is the dust-only 24\mi~monochromatic luminosity and 
$L_{H\alpha}$ is the observed \ha~luminosity. 

Therefore, we first restrict the variations of the three SSPs to their corresponding age 
ranges mentioned above in order to model the SEDs of various star-forming regions
across the merging disks in the Antennae galaxies.
Then we try to find the best-fit composite models with the $\chi^{2}$ minimization technique
\begin{equation}
\chi^2 = \sum_{\rm i=FUV}^{\it Ks} \left(\frac{F_{obs,i}-a F_{mod,i}}
{\sigma(F_{obs,i})}\right)^2 \label {eqn:eq2}
\end{equation} 
by minimizing Equation~\ref{eqn:eq2}.
In this way, we obtain the four 
 best-fit parameters, namely, the mass of the old (M$_{\rm o}$), intermediate 
(M$_{\rm i}$) and young (M$_{\rm y}$) populations, the extinction ratio ($\mu$) of 
intermediate/old to young populations. 
It should be stressed here that we first fix the extinction \av~ with Equation~1 prior to SED 
modeling, which affects mostly the young stellar populations. This makes our stellar population 
analysis presented in this paper much less affected by the degeneracy between stellar population 
and extinction. 

\subsection{Results of SED-fitting}

\subsubsection{Best-fit Parameters}

Our main fitting results are summarized in Table~\ref{table:1}. 
One should keep in mind that, we are actually comparing relatively clustered younger stellar populations with relatively extended older populations. Hence, the absolute values for the mass ratios between different populations should be sensitive to the size of photometric apertures and the average density of the underlying stellar populations. On that account we mainly focus on the relative changes of the mass ratios between different populations across the system. 

We notice that the SED-fitting quality for some {\it overlap} regions (i.e. regions 8, 9), 
is not as good as for most other regions  (Figure~\ref{fig:sed1}). The possible reason 
can be either the significant starlight contamination by nearby bright regions or the very faint 
nature (low S/N ratio) of these two regions at short wavebands. 
To check the significance of the starlight contamination 
by nearby bright regions, we also present the broadband SEDs for the four optically 
faint regions in the {\it overlap} region in Figure~\ref{fig:sed2}. 
It can be seen that the contamination by nearby bright regions does not 
change the overall SEDs shape for these regions. In fact, when we fit the SEDs (excluding 
the FUV/NUV data) extracted from the images at their original resolutions for these regions, 
our main conclusions for the {\it overlap} regions do not change. 
Table~\ref{table:2} lists the fitting results for the {\it overlap} regions using SEDs extracted 
from the images at their original resolutions. 
According to our best-fitting results for these regions, the best-fit ages of intermediate 
populations $\la$ 100 Myr, and the best-fitting ages of old stellar population $\geq$ 3~Gyr. 

\begin{table*}
\centering
\begin{minipage}{141mm}
\caption{SED-fitting Results.}
\label{table:1}
\begin{tabular*}{141mm}{@{\extracolsep{\fill}}ccccccccc@{}}
\hline\hline
\\
Reg. & $\chi^{2}_{r}$ & $\mu$ & M$_{\rm i}$/M$_{\rm o}$ & M$_{\rm y}$/M$_{\rm o}$ & M$_{\rm y}$/M$_{\rm i}$ & $\frac{\rm M_{y}/Age_{y}}{\rm M_{i}/Age_{i}}$ & $\frac{\rm M_{y}/Age_{y}}{\rm M_{o}/Age_{o}}$ & $\log$[Total Abs.](erg/s) \\
\hline
\multicolumn{9}{c}{NGC4039 Arms}\\
\hline
    1 ......& 1.1 &  0.1 & 0.003 & 7.0e-4 & 0.23  & 2.6 & 0.5 & 41.71 \\
\hline
\multicolumn{8}{c}{Southern Nucleus}\\
\hline
    2 ......& 2.2 & 0.9 & 0.03  & 6.0e-4 & 0.02  & 3.8 & 1.7 & 42.72 \\
\hline
\multicolumn{9}{c}{Overlap Regions}\\
\hline
    3 ......& 2.5 & 0.6 & 0.03       & 0.003    & 0.1     & 6.3     & 20.2 & 43.09 \\
    4 ......& 2.7 & 0.3 & 0.01       & 0.007    & 0.7     & 10.4    & 11.3 & 43.21 \\
    5 ......& 4.3 & 0.4 & 0.002      & 0.001    & 0.5     & 7.5     & 12.6 & 42.67 \\
    6 ......& 2.3 & 0.8 & 0.008      & 0.002    & 0.2     & 3.7     & 29.0 & 42.82 \\
    7 ......& 4.5 & 1.0 & 0.007      & 0.002    & 0.3     & 8.0     & 9.6  & 42.53 \\
    8 ......& 4.2 & 1.0 & 0.004      & 7.0e-4   & 0.2     & 11.1    & 6.1  & 42.61 \\
    9 ......& 5.9 & 0.8 & $<$1.0e-4  & 4.0e-4   & $>$4.0  & $>$40.0 & 3.0  & 42.46 \\
\hline
\multicolumn{9}{c}{Eastern Regions}\\
\hline
    10 ......& 3.0 & 0.9 & 0.01  & 5.0e-4 & 0.03   & 1.5 & 5.3 & 42.55 \\
    11 ......& 1.8 & 1.0 & 0.05  & 5.0e-4 & 0.01   & 0.2 & 2.1 & 42.46 \\
    12 ......& 1.9 & 0.9 & 0.05  & 4.0e-4 & 0.008  & 0.4 & 5.0 & 42.55 \\
    13 ......& 2.8 & 1.0 & 0.01  & 7.0e-4 & 0.07   & 1.3 & 3.4 & 42.60 \\
    14 ......& 2.1 & 0.3 & 0.01  & 6.0e-4 & 0.06   & 2.0 & 8.7 & 42.28 \\
    15 ......& 0.9 & 0.5 & 0.04  & 5.0e-4 & 0.01   & 1.3 & 4.3 & 42.36 \\
\hline
\multicolumn{9}{c}{Northern Regions}\\
\hline
    16 ......& 1.3 & 0.1 & 0.07   & 6.0e-4  & 0.008  & 1.5 & 1.7 & 41.57 \\
    17 ......& 1.5 & 0.2 & 0.1    & 6.0e-4  & 0.005  & 1.1 & 1.7 & 41.82 \\
    18 ......& 1.1 & 0.2 & 0.04   & 3.0e-4  & 0.007  & 0.7 & 4.3 & 41.91 \\
    19 ......& 1.0 & 0.3 & 0.05   & 3.0e-4  & 0.006  & 0.6 & 3.7 & 41.89 \\
    20 ......& 1.1 & 0.2 & 0.07   & 4.0e-4  & 0.006  & 0.6 & 3.4 & 41.62 \\
    21 ......& 1.2 & 0.1 & 0.04   & 4.0e-4  & 0.009  & 0.9 & 5.8 & 41.78 \\
    22 ......& 2.2 & 0.1 & 0.06   & 0.001   & 0.02   & 1.7 & 2.8 & 41.92 \\
\hline
\multicolumn{9}{c}{Western-Loop Regions}\\
\hline
    23 ......& 0.5 & 0.5 & 0.03   & 6.0e-4  & 0.02   & 0.3 & 4.4  & 42.42 \\
    24 ......& 2.3 & 0.7 & 0.09   & 4.0e-4  & 0.004  & 0.7 & 2.9  & 42.09 \\
    25 ......& 3.1 & 0.8 & 0.03   & 8.0e-4  & 0.02   & 0.8 & 11.6 & 42.46 \\
    26 ......& 1.9 & 0.7 & 0.2    & 0.007   & 0.04   & 0.5 & 50.0 & 42.72 \\
    27 ......& 3.1 & 1.0 & 0.2    & 0.006   & 0.03   & 0.4 & 29.6 & 42.72 \\
    28 ......& 3.8 & 1.0 & 0.2    & 0.002   & 0.01   & 1.3 & 17.4 & 42.30 \\
    29 ......& 3.1 & 1.0 & 0.2    & 0.003   & 0.02   & 1.8 & 3.0  & 42.26 \\
    30 ......& 3.0 & 1.0 & 0.1    & 0.002   & 0.02   & 0.6 & 3.2  & 42.43 \\
    31 ......& 2.5 & 0.7 & 0.03   & 7.0e-4  & 0.02   & 1.2 & 7.4  & 42.50 \\
    32 ......& 2.6 & 0.1 & 0.02   & 0.002   & 0.1    & 8.0 & 1.5  & 42.05 \\
\hline
\multicolumn{9}{c}{Circumnuclear Regions of NGC4038}\\
\hline
    33 ......& 1.1 & 0.8 & 0.03   & 4.0e-4   & 0.01   & 0.3 & 1.3 & 42.90 \\
\hline
\multicolumn{9}{c}{Northern Nucleus}\\
\hline
    34 ......& 1.4 & 0.6 & 0.01   & 7.0e-4  & 0.05   & 1.5 & 1.4 & 43.03 \\
\hline
\end{tabular*}

\medskip
$\chi^{2}_{r}$ is the reduced minimum $\chi^2$ value. 
$\mu$ is the extinction ratio of intermediate/old to young
populations. M$_{\rm y}$, M$_{\rm i}$ and M$_{\rm o}$ represent the mass
of young, intermediate and old populations respectively. 
$\frac{\rm M_{y}/Age_{y}}{\rm M_{i}/Age_{i}}$ and $\frac{\rm M_{y}/Age_{o}}{\rm M_{i}/Age_{o}}$ 
represent the corresponding age-averaged mass fraction ratios. 
The total starlight absorption obtained from the models is listed in the last column.
\end{minipage}
\end{table*}

\begin{table*}
\centering
\begin{minipage}{141mm}
\caption{Fitting Results for the SEDs (without FUV/NUV) extracted from the images at their original 
resolutions. Only results for some of the {\it overlap} regions which may have significant mutual light 
contamination are listed.}
\centering
\label{table:2}
\begin{tabular*}{141mm}{@{\extracolsep{\fill}}ccccccccc}
\hline\hline
\multicolumn{9}{c}{Overlap Regions}\\
\hline
Reg. & $\chi^{2}_{r}$ & $\mu$ & M$_{\rm i}$/M$_{\rm o}$ & M$_{\rm y}$/M$_{\rm o}$ & M$_{\rm y}$/M$_{\rm i}$ 
&  $\frac{\rm M_{y}/Age_{y}}{\rm M_{i}/Age_{i}}$ & $\frac{\rm M_{y}/Age_{y}}{\rm M_{o}/Age_{o}}$ & $\log$[Total Abs.](erg/s) \\
\hline
    3 ......& 1.0  & 0.5  & 0.04      & 0.01    & 0.3    & 5.0  & 40.0  & 42.97 \\
    4 ......& 0.3  & 0.3  & 0.01      & 0.05    & 4.5    & 9.4  & 100.1 & 43.50 \\
    5 ......& 0.7  & 0.8  & 0.004     & 0.005   & 1.3    & 4.7  & 13.5  & 43.11 \\
    6 ......& 1.9  & 1.0  & 0.05      & 0.01    & 0.2    & 1.5  & 73.0  & 42.94 \\
    7 ......& 1.8  & 0.3  & 0.02      & 0.004   & 0.2    & 2.9  & 42.5  & 42.54 \\
    8 ......& 0.9  & 1.0  & 0.04      & 0.002   & 0.05   & 0.6  & 8.8   & 42.82 \\
    9 ......& 1.0  & 0.7  & 0.002     & 0.001   & 0.5    & 5.0  & 4.2   & 42.52 \\
\hline
\end{tabular*}
\end{minipage}
\end{table*}

For the mass ratios of intermediate to old populations M$_{\rm i}$/M$_{\rm o}$, 
the {\it overlap} regions have overall smaller values compared to 
other regions, suggesting the {\it overlap} star-forming regions are very young. 
Northern edge of the {\it overlap} regions, i.e. region 9, for example, has the 
lowest M$_{\rm i}$/M$_{\rm o}$. Whereas the {\it western-loop} regions of NGC~4038 have overall the 
largest M$_{\rm i}$/M$_{\rm o}$ across the whole system, and they are the brighest regions in UV. 
The {\it western-loop} regions host most of the intermediate populations. 
The ratios of most other regions across the whole system fall in between the {\it overlap} 
and {\it western-loop} regions. 

For the mass ratios of young to old populations M$_{\rm y}$/M$_{\rm o}$, 
the {\it overlap} regions and some {\it western-loop} regions have M$_{\rm y}$/M$_{\rm o}$ 
larger than all the other regions. 
The similar contrast could also been seen from the age-averaged mass fraction ratios of 
young to old populations ($\frac{\rm M_{y}/Age_{y}}{\rm M_{o}/Age_{o}}$). 
The {\it overlap} and {\it western-loop} regions host most of the young populations, and 
they are the most intense current star-forming sites across the whole system. 

For the mass ratios of young to intermediate populations M$_{\rm y}$/M$_{\rm i}$, 
the {\it overlap} regions hold the largest values, while the {\it northern} 
regions have overall small M$_{\rm y}$/M$_{\rm i}$, conforming with their weak 
{\it current} star formation activities. 
We also point out here one interesting finding. That is, across the $overlap$ 
regions, the northwestern edge, namely region 9, and the southeastern edge, region 4, 
have the M$_{\rm y}$/M$_{\rm i}$ ratios significantly larger than the central regions, 
e.g. region 6. We note that the trend still exist after we normalize the mass fraction 
ratios with their corresponding best-fit ages ($\frac{\rm M_{y}/Age_{y}}{\rm M_{i}/Age_{i}}$). 

\begin{figure}
\centering
\includegraphics[scale=0.37]{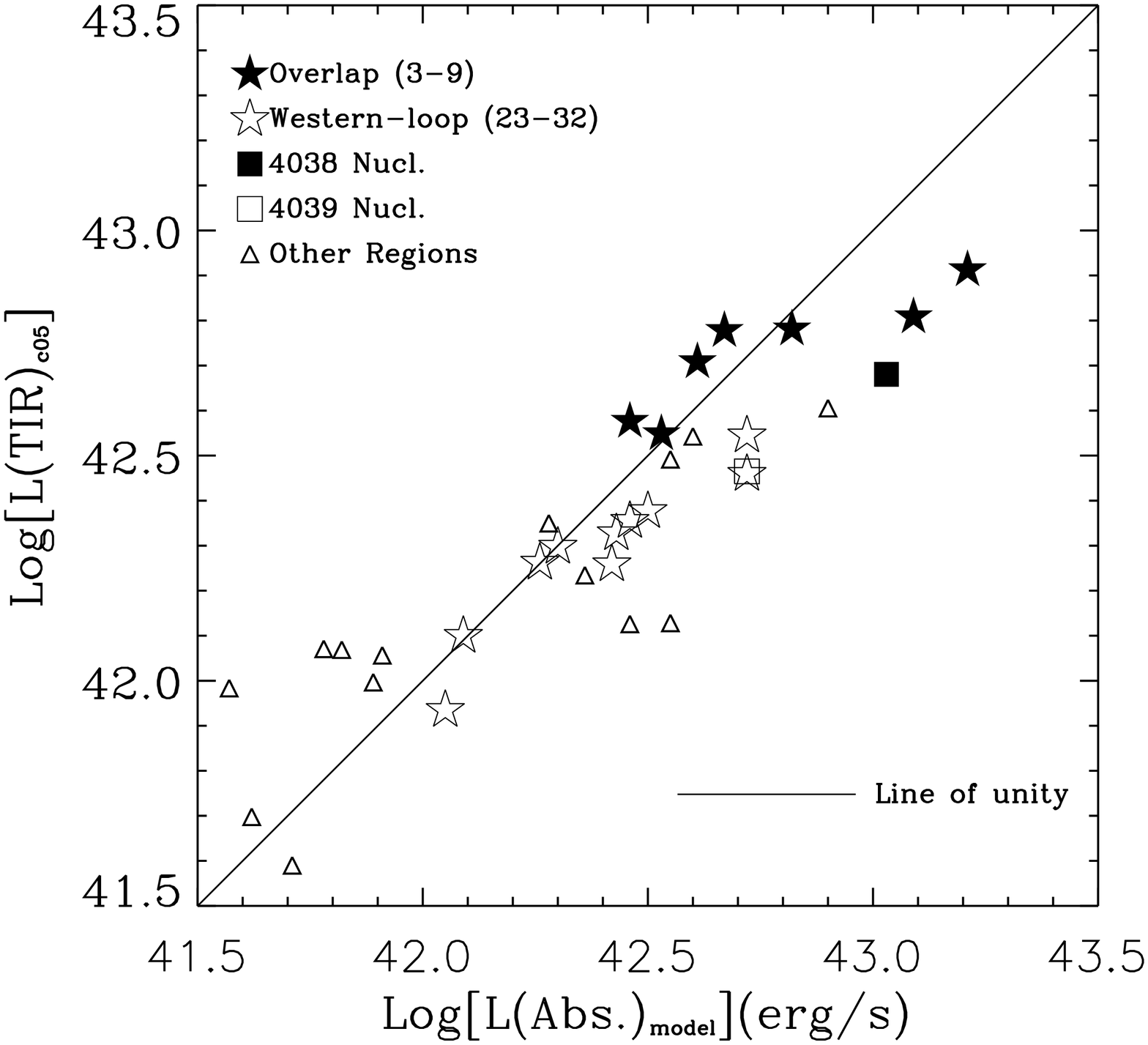}
\caption{Total starlight absorption obtained from the SED-fitting models is compared 
with total infrared luminosity estimated from dust-only 24 $\mu m$ and 
8 $\mu m$ using the relation calibrated in NGC~5194 by \citet{calzetti2005} (hereafter TIR$_{c05}$).
The line with slope of unity and null intercept is also plotted ({\it solid line}). 
}\label{fig:ira}
\end{figure}

In the last column of Table~\ref{table:1}, we also list the corresponding 
total starlight absorbed by dust which should be equal to the total infrared (TIR) 
dust emission. \citet{calzetti2005} exploited a relationship between 24\mi/TIR and 
8/24\mi~for the star-forming regions in NGC~5194. 
Considering the similar global properties between the Antennae 
and NGC~5194, such as similiar PAH abundances (the PAH index $\sim$ 4.5 according to 
\citet{Draine2007} dust emission models), comparable TIR \citep{Sanders2003}
and total molecular gas surface density \citep{Wilson2003}, we compare the total 
starlight absorption from our models with the TIR estimated from dust-only 8\mi~and 24\mi~
using the relation (equation 1 in \citealt{calzetti2005}) exploited for the star-forming 
regions of NGC~5194. The comparison is shown in Figure~\ref{fig:ira}.
Obviously, almost all our selected star-forming regions in Arp~244 generally follow the same
relation derived in the star-forming regions of NGC~5194. 
Considering the general consistency,  
in the following sections, we take the total starlight absorption from our SED-fitting models 
as the best estimate of TIR for our selected star-forming regions.  

\subsubsection{Star-Forming Regions of Arp~244 on the IRX-UV Plane}

\begin{figure}
\centering
\includegraphics[scale=0.38]{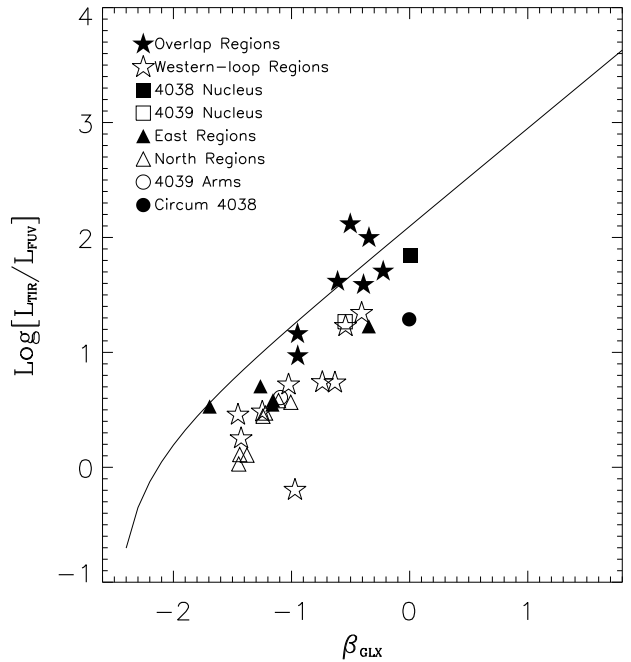}
\caption{Ratio of TIR (set to total starlight absorption from models) to FUV luminosity as
a function of equivalent ultraviolet spectral slope (the IRX-UV diagram).
Data for different star-forming regions within the Antennae are plotted with different symbols.
The solid line respresents the mean IRX-UV relation for 50 starburst
galaxies fitted by \citet{kong2004}.} \label{fig:irx}
\end{figure}

\citet{kong2004} found that, while the dust extinction is the main factor driving the
strong correlation between $L_{\rm TIR}/L_{\rm FUV}$ (hereafter IRX) and UV
spectral slope, it is the star formation history that affects the degree of deviation of a
star-forming galaxy from the locus of starbursts in the IRX-UV color indices' plane.
It is surely meaningful to plot these indices for the star-forming regions of the Antennae 
on the IRX-UV color plane.
Figure~\ref{fig:irx} shows the ratio of $L_{\rm TIR}/L_{\rm FUV}$, where TIR
is set to be equal to the total starlight absorption obtained from our models,
as a function of $\beta_{\rm GLX}$ ($\beta_{\rm GLX} = \frac{\log F_{\rm FUV}-\log F_{\rm NUV}} 
{\log \lambda_{\rm FUV}-\log \lambda_{\rm NUV}}$, see Kong et al. 2004).
The solid line represents the tight correlation for starburst galaxies \citep{kong2004}.
Obviously, except the {\it overlap} regions, almost all the star-forming regions 
in Arp~244 lie below the locus followed by starbursts. The three regions that lie above the starburst 
locus are region 4, 5 and 9. Hence, it can be stated here that except for some localized starburst 
sites, mainly 
in the {\it overlap} regions, most regions across the whole system are forming stars at a level weaker
than that of the starbursts. This is also consistent with our further analysis on star formation 
histories.  

\begin{figure}
\centering
\includegraphics[scale=0.38]{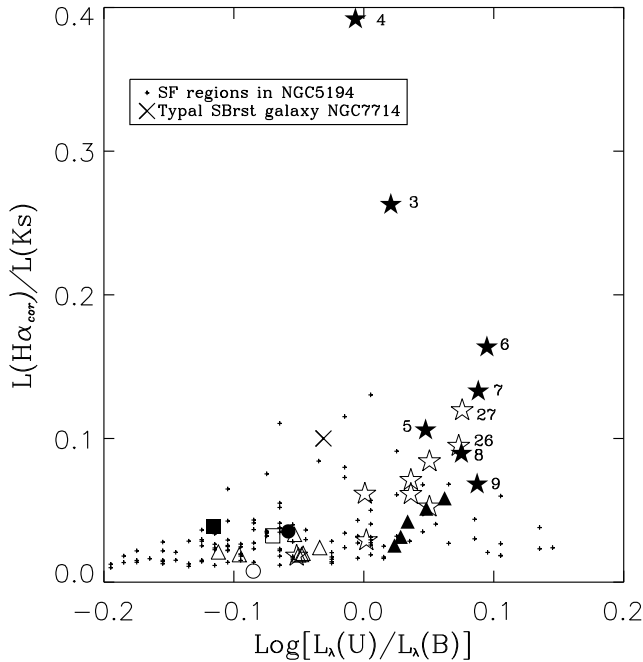}
\caption{Ratio of extinction-corrected (using 24\mi~) $L(H\alpha_{corr})$ to $L(Ks)$ vs. 
$\log[L_\lambda(U)/L_\lambda(B)]$. The symbols are as Figure~\ref{fig:irx}.
The star-forming regions selected by \citet{calzetti2005} in the quiescently star-forming galaxy 
NGC~5194 are shown by small {\it pluses}. The archetypal starburst galaxy NGC~7714 as a whole is represented by {\it large cross}. 
Obviously, just like the majority of star-forming regions in NGC~5194, most regions (except some of the {\it overlap} and {\it western-loop} regions) in Arp~244 host a significant fraction of 
older stellar populations and have relatively lower current star formation strength compared to 
typical starbursts. 
See the text for further explanation.} \label{fig:hak}
\end{figure}

Figure~\ref{fig:hak}  plots the ratios of $L(H\alpha_{corr})/L(Ks)$  vs. 
$L_{\lambda}({U})/L_{\lambda}(B)$. 
Star-forming regions in NGC~5194 selected (mainly as 24\mi~peaks) by \citet{calzetti2005} are shown as small {\it pluses}. Also plotted is the archetypical starburst galaxy NGC~7714 as a whole ({\it large cross}). The modest total star formation rate ($\sim$ 3.4 \msun yr$^{-1}$) and star formation intensity ($\sim$ 0.015 \msun yr$^{-1}$Kpc$^{-2}$) of NGC~5194 place it among the quiescently  star-forming galaxies, although it hosts a LINER-type nucleus. \citet{calzetti2005} found that the star-forming regions in NGC~5194 have properties quite similar to that of the normal star-forming galaxies, rather than that of starbursts. 
Assuming $L_{\lambda}(U)/L_{\lambda}(B)$ and $L(H\alpha_{corr})/L(Ks)$ roughly represent 
the ratios of recent-to-past and current-to-past star formation strengths, respectively, 
Figure~\ref{fig:hak} reveals similar results as that of the IRX-UV diagram. 
Namely, almost all the star forming regions of Arp~244 have current star formation strength 
comparable with star-forming regions in quiescently star-forming galaxies (low $L(H\alpha_{corr})/L(Ks)$), 
except for some of the {\it overlap} and {\it western-loop} regions. 

\subsubsection{Emission Excess at IRAC 3.6\mi~and 4.5\mi}
Since we have done relatively sophisticated SED-fitting for these star-forming regions, 
we could estimate the hot dust emission at IRAC 3.6\mi~and 4.5\mi~which are 
often assumed to be dominated by stellar photospheric emission in literatures.  
With the average of 3.6\mi~and 4.5\mi~fluxes as the underlying stellar 
continuum level, \citet{Wang2004} concluded that most of the emission ($>$90\%) 
in these two bands comes from direct stellar contribution for the whole system.
From our stellar SED-fitting, we could easily
see (Figures~\ref{fig:sed1} and \ref{fig:sed2}) the prominent emission 
excess at 3.6\mi~( $\sim$ 10\% - 65\%) and 4.5\mi~( $\sim$ 10\% - 80\%) 
for most of these star-forming regions relative to the stellar  
(plus nebular continuum) emission, especially for the {\it overlap} regions. 
The significant emission excess at 3.6\mi~and 4.5\mi~brings out the caution 
needed in using them as representation for the underlying stellar mass, particularly 
when studying the active star-forming regions. 
Since we have largely accounted for hydrogen/helium recombination emission in our SED-fitting, 
according to the existing observations of spectra in these two bands of compact H$_{~\rm II}$ regions  
(e.g. \citealt{martin2000}; \citealt{peeters2002}),
the other possible main mechanisms of the emission excess could be vibrationally excited 
molecular hydrogen emission, atomic fine-structure lines, or very small hot dust (e.g. 3.3\mi~PAH feature) emission. 

For region 4, the MIR ``{\it hotspot}'', the excess emission components 
contribute $\ge$ 65\% and 80\%
\footnote{Considering the hot dust emission contribution to Ks band, our estimation 
should be lower limits.
}
of the total emission at 3.6\mi~and 4.5\mi~respectively.  
As is already pointed out by \citet{Wang2004}, the extremely high obscuration 
of region 4 must be responsible for the large excess to some extent. 
The high-quality MIR spectrum of this region obtained from IRS on board {\it Spitzer} 
is characterized by prominent PAHs and very hot VSGs emission \citep{brandl2009}. 
Given the prominent PAH feature emission of region 4, the exceptionally low excess 
emission flux ratio of 3.6\mi/4.5\mi~($\sim$ 0.9) can not be accounted for by the pure 
dust emission models of \citet{Draine2007}. 
Furthermore, the {\em K}-band spectra \citep{Gilbert2000}  in the observations of the 
compact star clusters coincident with the MIR ``{\it hotspot}'' are characterized 
by prominent nebular and fluorescent H$_2$ emission with slightly rising 
continuum toward the red (due to hot dust emission). 
The fact that region 4 has very flat radio spectrum \citep{Neff2000}
and prominent PAH feature emission excludes the shock origin of the possible molecular/atomic emission.
Thus, the fine-structure lines from heavy elements, the molecular lines from FUV 
fluorescence in dense photodissociation regions (PDRs) and hot dust (including PAHs) emission may all 
have significant contribution to the emission excess at IRAC 3.6\mi~and 4.5\mi.

Interestingly, the total 4.5\mi~flux of region 4 is even $\sim$ 20\% higher 
than that of the southern nucleus and nearly equal to that of northern nucleus.
In short, the excess emission (either from hot dust or molecule/ionized atoms) 
at IRAC 4.5\mi~bands have nonnegligible ($>$ 18\% 
\footnote{This is obtained by the sum of flux of excess emission for all the star-forming 
regions selected in this work divided by the total flux for the whole system. 
So here our estimation is a lower limit for the whole system.
}
) contribution to the total flux for the whole system. 

\section{Discussion} \label{sec:dis}
\subsection{\it Star Formation Histories Across the Whole System}
Since we have the mass ratios between the three stellar populations for these 
star-forming regions, using the best-fit ages of the three populations for each region, 
we could easily derive the ratios between the age-averaged star formation rate (SFR) 
for the three (i.e. current, recent and past) star formation epoches related to our 
three populations (Table~\ref{table:1} and 2). 
We find that,  
except for some of the {\it overlap} regions and the {\it western-loop} regions 
of NGC~4038, which have the ratios of {\it current}-to-{\it past} age-averaged SFR 
(defined approximately as $\frac{\rm M_{y}/Age_{y}}{\rm M_{o}/Age_{o}}$)$\ga10$, 
almost all the other star-forming regions have the ratios $\ga1$. Specifically, 
the regions which have the ratios $\ga10$ are 3, 4, 5, 6, 25, 26, 27, and 28. 
Hence, it can be said that, some of the {\it overlap} and {\it western-loop} regions 
are experiencing bursts of star formation, whereas most of the other regions are
forming stars at a moderate level comparable with normal star-forming galaxies.
The location of these star-forming regions on the IRX-UV plane (Figure~\ref{fig:irx})
is also consistent with the fact that most of the regions are at a modest star formation intensity. 
For the ratios of {\it current}-to-{\it recent} age-averaged SFR (defined as 
$\frac{\rm M_{y}/Age_{y}}{\rm M_{i}/Age_{i}}$) (see Table~\ref{table:1}), 
the {\it overlap} regions have the ratios $\ga 3$, whereas almost all the other regions 
have the ratios $\sim 1$. The high $\frac{\rm M_{y}/Age_{y}}{\rm M_{i}/Age_{i}}$ for the two 
nuclei may be attributed to their surrounding star-forming regions. 
Hence, comparing with the {\it overlap} regions, 
the star formation of most regions across the whole system is not only modest (except 
the {\it western-loop} regions) but also continuous during the recent tens of million years 
(typical best-fit ages of the intermediate stellar populations for these regions).
Namely, while both the {\it overlap} regions, and the {\it western-loop} regions are the most 
intense {\it current} star-forming sites across the whole system, the {\it overlap} regions 
are now experiencing much more violent enhancement of star formation compared to all the other regions. 

In fact, the star formation histories for these star-forming regions could also be 
simply probed through the ratios of $L_{\lambda}(U)/L_{\lambda}(B)$ and 
$L(H\alpha_{corr})/L(Ks)$, which roughly represent ratios of the recent-to-past star 
formation strength and the current-to-past star formation strength, respectively. 
Figure~\ref{fig:hak} clearly shows that, except for the {\it overlap} and {\it western-loop} 
regions, most regions are forming stars at moderate level (low $L(H\alpha_{corr})/L(Ks)$) 
comparable with star-forming regions in quiescently star-forming galaxies. 
And meanwhile these regions host a remarkable fraction of aging stellar populations 
(small $L_{\lambda}(U)/L_{\lambda}(B)$), just like the star-forming regions in NGC~5194. 
The overall larger ratios $L_{\lambda}(U)/L_{\lambda}(B)$ for the {\it overlap}, 
the {\it western-loop} and the {\it eastern} (to a lesser extent) regions compared with 
the star-forming regions in quiescently star-forming galaxies indicate that these regions may 
have just experienced a period of intense starburst in the recent few hundred million years. 
It is also notable of the much higher ratios of current-to-past star formation strength for 
some of the {\it overlap} and {\it western-loop} regions. \citet{Wang2004} also got similar 
results from the flux ratio of the dust-only IRAC 8\mi~and the underlying stellar continuum. 

\subsection{\it The 20~cm-to-CO ratio map as star formation efficiency map}

Taking advantage of the tight correlation between FIR and radio continuum (\citealt{condon1992}; \citealt{yun2001}; \citealt{bell2003}), which appears to be valid at least 
on kiloparsec scales in galaxies (\citealt{lu1996}), \citet{Gao2001} constructed a star formation 
efficiency map using the 20~cm-to-CO ratio. Overall, our results of population analysis are consistent 
with their star formation efficiency map, viz, except some localized starburst sites mainly in the 
{\it overlap} and {\it western-loop} regios, most regions across the whole system are forming stars 
at a quite moderate level comparable with normal star-forming galaxies. 
This verifies the practice of using radio continuum as an indicator of the star formation and the 
radio-to-CO ratio maps as representation of star formation efficiency maps in most cases. 

Nevertheless, some exceptions do exist, like for some overlap regions of galaxy pairs in high-speed 
collision. For instance, in the Taffy galaxy a large portion of the synchrotron radio emission may be 
related to gas collision shocks, rather than supernovae remnants (SNRs, most likely related to recent 
star formation) shocks (\citealt{gao2003}; \citealt{zhu2007}). In the {\it overlap} regions of the 
Antennae, \citet{Gao2001} found that the radio-to-CO ratios progressively increase from the 
southeastern side to the northwestern edge across the {\it overlap} regions. 
However, we note that, the lower radio-to-CO ratio of the southeastern region (4) compared to 
other {\it overlap} regions may be, rather than due to lower star formation 
efficiency, assigned to other various causes.
First, the very flat radio continuum suggests that the 20~cm continuum in region 4 
, rather than being dominated by synchrotron emission from (maybe) SNRs shocks like other 
{\it overlap} regions, is primarily thermal free-free emission from young, compact H$_{~\rm II}$ regions. 
This means that, in such violent overlap starburst envirionment, large numbers of supernovae 
events associated with current star formation epoch have not happened. 
Second, studies on the local IR/radio correlation within nearby galaxies 
(e.g. \citealt{hughes2006}; \citealt{murphy2006}) demonstrate the weak trend of increasing 
IR/radio ratio with increasing IR luminosity within individual galaxy. Both our TIR 
estimates (equal to the total starlight absorption) and recent IR estimate based on the 
15\mi~and 30\mi~continuum fluxes (\citealt{brandl2009}) suggest that region 4 has the 
largest IR luminosity across the whole system. 
Finally, the very hot dust emission, as we have shown above, for the southeastern 
side (i.e. region 4) implies that most energy there may emit in the MIR regime, 
which again is different from most of the other regions. 
Therefore, unlike the usually continuous star formation in a relatively long timespan 
($\ge 10^{8}$ yr), in the violent galaxies interaction regions,
both the strong shocks as a result of cloud-cloud collisions and the strong variations of 
star formation rate in short timescales (tens of Myr) may all make the 20cm-to-CO ratio 
fails to be an efficient star formation efficiency indicator.

\subsection{\it Sequential Star Formation Paths in the {\it Overlap} Regions}

Across the {\it overlap} regions (Table~\ref{table:1} and 2), 
the northwestern edge and the southeastern edge have both higher mass ratios of young to intermediate 
populations M$_{\rm y}$/M$_{\rm i}$ (i.e. regions 9 and 4) and higher correspondingly age-averaged 
SFR ratios $\frac{\rm M_{y}/Age_{y}}{\rm M_{i}/Age_{i}}$ than do the central regions (e.g. regions 6). 
We like to interpret the trends as two sequential star formation paths. 
One is from the central regions (i.e. regions 6) to the {\it southern} edge (e.g. region 4), 
and the other is from the central to the {\it northwestern} edge (e.g. region 9).

To check if there are any trends for the spatial distributions of the star clusters 
across the {\it overlap} regions, we refer to the cluster distribution maps derived 
by \citet{zhang2001}. We find a slightly proportionally deficit of clusters with ages of tens of 
Myr for both the northwestern (corresponding to our region 9) and the southern edges compared to the central regions (e.g. region 6). 
Recently, \citet{Mengel2005} also obtained the spatial distributions for clusters with 
Br$\gamma$-determined and CO-index-determined ages. 
Interestingly, the distributions also show the relative deficit of intermediate-age 
($\sim$ 10 Myr) clusters for both the northwestern and southern edges, although 
the sample size and the probed age range are all small compared to that of \citet{zhang2001}. 
The cluster distributions are in agreement with the sequential star formation paths 
mentioned above, i.e. both the {\it northwestern} and {\it southern} regions are just beginning 
their recent star formation episode.   

\citet{Jog1992} proposed a physical mechanism to explain the origin of the enhanced 
star formation occurring in situ in the overlapping regions of a pair of colliding 
galaxies like the Antennae. 
In their model, following a collision between galaxies, the H I cloud-cloud 
collisions from the two galaxies lead to the formation of hot, ionized, high-pressure 
remnant gas that compresses the outer layers of preexisting GMCs in the overlapping regions. 
This makes the GMCs shells become gravitationally unstable, which triggers a starburst in the 
initially barely stable GMCs. 
Although generally H I cloud-cloud collisons should be more efficient than that 
of the GMCs due to the much smaller mean free path of an H I cloud than GMC, the 
huge concentration of molecular gas in the {\it overlap} regions of Arp~244
(\citealt{Wilson2000}; \citealt{Gao2001}) may make the GMCs collisions also possible. 

Based upon this model, across the {\it overlap} regions of Arp~244, 
the sequential star formation paths we have found can be explained naturally. 
As the two colliding disks begin to interpenetrate each other, the regions that 
overlap first, e.g. region 6, may have more of the gravitationally unstable GMCs 
layers formed first, leading to in situ starburst first there. 
As the colliding/merging proceeds and the overlapping zone between 
the two colliding disks bulks up, cloud collisions from the two 
colliding disks spread to more and more regions, i.e. 
both the northwest (e.g. to regions 8, 9) and southeast (e.g. to regions 5, 4) 
of the {\it overlap} regions. This leads to progressively lagging starbursts triggered 
by the radiative shock compressions toward both the northwest and southeast 
directions of the {\it overlap} regions. 
In short, the identified sequential star formation paths could be the imprints 
of the interpenetrating process of the two colliding galaxy disks.

Both kinematic analysis \citep{Hibbard2001} and numerical simulations 
(\citealt{Toomre1972}; \citealt{barnes1988}) suggest that the two galaxies of the 
Antennae system begun their first close encounter several hundred million years ago, 
and the two colliding galaxies may have passed their first close encounter \citep{Mihos1993}. 
The first peak of large-scale starburst phase may have just passed in the Antennae 
\citep{Mihos1996}, which can be evidenced by the moderate star formation strengths
shown in Figs. 5 \& 6, consistent with the star formation efficiency map across 
the whole system \citep{Gao2001}. 
While the moderate, continuous (recent) star formation for most regions is consistent with this scenario, the currently violent enhancement of star formation and the sequential star formation paths for the {\it overlap} regions may suggest that now the two colliding galaxies are just launching their second close encounter. 

After analysing H$_{2}~\nu = 0-0~S(3)~\lambda = 9.66~\mu m$
line emission obtained by ISOCAM CVF,
\citet{Haas2005} found that both the southwestern and northwestern edges of
the {\it overlap} regions, which are very close to the star-forming region 5 and
regions 9, respectively, have exceptionally high $L(H_{2})/L(FIR)$ ratios that
exceed that of all other known galaxies. But the absolute current/recent star formation  
there (especially the northwestern edge) are very weak, just as we find from population analysis 
(see Table~\ref{table:1} and 2). They suggest that the high H$_{2}$ emission 
there should be excited by pre-starburst shocks which are caused by cloud-cloud collisions. 
The low mass fractions of intermediate populations and the very high (age-averaged) mass
ratios of young to intermediate populations for the two edges, especially the northwestern
edge, indicate there indeed are very young star-forming sites, which are most probablely to
be triggered by pre-starburst shocks following the second close encounter between the two galaxies.
However, recent high-quality observations from {\it Spitzer} (\citealt{brandl2009}) detected 
about five times less integrated H$_{2}$~S(3) line flux than does ISOCAM CVF, and didn't find 
the previously claimed strong H$_{2}$ emission peak in the northern {\it overlap} zone. 
These new observations cast doubt on the pre-starburst shock origin of the H$_{2}$ emission in the {\it overlap} regions. 

\section{Summary} \label{sec:sumr}
To summarise, taking advantage of the availability of 
multiwavelengh imagery from FUV to 24~\mi~from \galex,~\hst,~\mass~and \spitzer~
in both high resolution and high sensitivity: 
\begin{itemize}
\item We compare the broadband SEDs of star-forming regions selected as 24\mi~peaks 
across the whole merging disks of the Antennae galaxies, which provides us a 
basis to comprehend the complete picture of star formation histories. 
The large ratios of 24\mi/8\mi~for the {\it overlap} regions and 
the blue, strong UV emission for the {\it western-loop} regions demonstrate 
that currently they are the most intense star-forming sites, although the {\it western-loop} 
regions are at a relatively later star formation stage compared to the {\it overlap} regions.  
Most of the other regions, which have redder {\it UB} color, weaker IR dust emission and UV emission, 
across the whole system are forming stars at a quite moderate level during the past $\sim$ 100 Myr. 
\item We roughly constrain the star formation histories of these active star-forming 
regions across the whole system, with the degeneracy between stellar population and extinction broken, 
by including 24\mi~dust emission into population analysis.
Compared with other regions, the {\it overlap} regions 
are now experiencing much more violent enhancement of star formation, 
although both the {\it overlap} and the {\it western-loop} regions are the most
intense {\it current} star-forming sites across the whole system. 
Our analysis is in general agreement with the findings of \citet{Gao2001},
i.e. except for some localized violent starbursts confined mainly in the {\it overlap}
regions and the {\it western-loop} regions of NGC~4038, 
the bulk of star formation is at a moderate level comparable to that 
of star-forming regions in quiescently star-forming galaxies.
\item We suggest two sequential star formation paths across the famous {\it overlap}
regions, which may reflect the (second) interpenetrating process of the
second passage between the two colliding galaxy disks.
We also suggest that the recent star formation of both the northern and southern edges of 
the {\it overlap} zone might be just triggered by pre-starburst shocks.
\item We report the nonnegligible ($>$ 18\%) excess emission contribution to 
the total IRAC 4.5\mi~for the whole system. The well-known brightest MIR ``{\it 
hotspot}'' in the {\it overlap} regions has total 4.5\mi~emission ($\ge$ 80\% 
excess emission) even higher (by $\sim$ 20\%) than that of 
the nuclear region of NGC~4039 and nearly equals to that of the nuclear region of 
northern galaxy NGC~4038. The unusually low ratio of 3.6\mi/4.5\mi~
implies that, in addition to hot dust emission, other emission mechanisms, such as 
atomic fine-structure lines and vibrationally excited molecular hydrogen lines from 
dense PDRs, have significant contribution to IRAC~4.5\mi. 
\end{itemize}

\section*{Acknowledgements}

We thank the anonymous referee who provided extremely useful comments that resulted in
substantial improvements to the presentation of the paper.
We are grateful to Y. H. Zhao and X. Z. Zheng for useful discussions and 
suggestions that helped improve the presentation of this paper. 
Research for this project is supported by NSF of China (Distinguished 
Young Scholars \#10425313, grants \#10833006 \& \#10621303), 
Chinese Academy of Sciences' Hundred Talent Program, and 973 project of 
the Ministry of Science and Technology of China (grant \#2007CB815406). 
XK is supported by the National Natural Science Foundation of 
China (NSFC, Nos. 10633020, and 10873012), the Knowledge Innovation 
Program of the Chinese Academy of Sciences (No. KJCX2-YW-T05), and 
National Basic Research Program of China (973 Program; No. 
2007CB815404).
This research has made use of the NED which is operated by the JPL.  
Archival data from \galex,~\hst,~\mass~and \spitzer~have also been 
used for the research.

\label{lastpage}

\begin{thebibliography}{88}

\bibitem[\protect\citeauthoryear{Alonso-Herrero et al.}{2006}]%
          {alonso2006} Alonso-Herrero A., Rieke G. H., Rieke M. J., Colina L., 
                       Perez-Gonzalez P. G., Ryder S. D., 2006, ApJ, 650, 835
\bibitem[\protect\citeauthoryear{Anders \& Fritze-v. Alvensleben}{2003}]%
          {Anders2003} Anders P., \& Fritze-v. Alvensleben U., 2003, A\&A, 401, 1063
\bibitem[\protect\citeauthoryear{Barnes}{1988}]%
          {barnes1988} Barnes J. E., 1988, ApJ, 331, 699
\bibitem[\protect\citeauthoryear{Bastian et al.}{2006}]%
          {bastian2006} Bastian N., Emsellem E., Kissler-Patig M., Maraston C., 2006, A\&A, 445, 471
\bibitem[\protect\citeauthoryear{Bastian et al.}{2009}]%
          {bastian2009} Bastian N., Trancho G., Konstantopoulos L. S.. Miller B. W., 2009, preprint (astro-ph/09062210)
\bibitem[\protect\citeauthoryear{Bell}{2003}]%
          {bell2003} Bell E. F., 2003, ApJ, 586, 794
\bibitem[\protect\citeauthoryear{Brandl et al.}{2009}]%
          {brandl2009} Brandl B. R., Snijders L., den Brok M. et al., 2009, 
                       preprint (astro-ph/09051058)
\bibitem[\protect\citeauthoryear{Calzetti, Kinney \& Storchi-Bergmann}{1994}]%
          {calzetti1994} Calzetti D., Kinney A. L., Storchi-Bergmann T., 1994, ApJ, 429, 582
\bibitem[\protect\citeauthoryear{Calzetti et al.}{2000}]%
          {calzetti2000} Calzetti D., Armus L., Bohlin R. C., Kinney A. L., Koornneef J., 
                         Storchi-Bergmann T., 2000, ApJ, 533, 682
\bibitem[\protect\citeauthoryear{Calzetti et al.}{2005}]%
          {calzetti2005} Calzetti D., Kennicutt R. C., Bianchi L. et al., 2005, ApJ, 633, 871 
\bibitem[\protect\citeauthoryear{Calzetti et al.}{2007}]%
          {calzetti2007} Calzetti D., Kennicutt R. C., Engelbracht C. W. et al., 2007, ApJ, 666, 870
\bibitem[\protect\citeauthoryear{Cardelli, Clayton \& Mathis}{1989}]%
          {cardelli1989} Cardelli J. A., Clayton G. C., Mathis J. S., 1989, ApJ, 345, 245
\bibitem[\protect\citeauthoryear{Charlot \& Fall}{2000}]%
          {Charlot2000} Charlot S., Fall S. M., 2000, ApJ, 539, 718
\bibitem[\protect\citeauthoryear{Condon}{1992}]%
          {condon1992} Condon J. J., 1992, ARA\&A, 30, 575
\bibitem[\protect\citeauthoryear{Conselice et al.}{2003}]%
          {conselice2003a} Conselice C. J., Bershady M. A., Dickinson M., Papovich C., 2003, AJ, 126, 1183
\bibitem[\protect\citeauthoryear{Conselice, Chapman \& Windhorst}{2003}]%
          {conselice2003b} Conselice C. J., Chapman S. C., Windhorst R. A., 2003, ApJ, 596, L5
\bibitem[\protect\citeauthoryear{Conselice, Yang  \& Bluck}{2009}]%
          {conselice2009} Conselice C. J., Yang C., Bluck A. F. L., 2009, MNRAS, 394, 1956
\bibitem[\protect\citeauthoryear{de Ravel et al.}{2008}]%
          {deravel2008} de Ravel L., Le Fevre O., Tresse L. et al., 2008, A\&A, 498, 379
\bibitem[\protect\citeauthoryear{Draine \& Li}{2007}]%
          {Draine2007} Draine B. T., Li A., 2007, ApJ, 657, 810
\bibitem[\protect\citeauthoryear{Elbaz \& Cesarsky}{2003}]%
          {Elbaz2003} Elbaz D. M., Cesarsky C. J., 2003, Science, 300, 270
\bibitem[\protect\citeauthoryear{Fabbiano, Zezas \& Murray}{2001}]%
          {fabbiano2001} Fabbiano G., Zezas A., Murray S. S., 2001, ApJ, 554, 1035
\bibitem[\protect\citeauthoryear{Fabbiano et al.}{2003}]%
          {fabbiano2003} Fabbiano G., Krauss M., Zezas A., Rots A., Neff S., 2003, ApJ, 598, 272
\bibitem[\protect\citeauthoryear{Fabbiano et al.}{2004}]%
          {fabbiano2004} Fabbiano G., Baldi A., King A. R., Ponman T. J., Raymond J., Read A., 
          Rots A., Schweizer F., Zezas A., 2004, ApJ, 605, 21
\bibitem[\protect\citeauthoryear{Fall, Chandar \& Whitmore}{2005}]%
          {fall2005} Fall S. M., Chandar R., Whitmore B. C., 2005, ApJ, 631, 133
\bibitem[\protect\citeauthoryear{Fazio et al.}{2004}]%
          {Fazio2004} Fazio G. G., Hora J. L., Allen L. E. et al., 2004, ApJS, 154, 10
\bibitem[\protect\citeauthoryear{Gao et al.}{2001}]%
          {Gao2001} Gao Y., Lo K. Y., Lee S. -W., Lee T. -H., 2001, ApJ, 548, 172
\bibitem[\protect\citeauthoryear{Gao et al.}{2003}]%
          {gao2003} Gao Y., Zhu M., Seaquist E. R., 2003, AJ, 126, 2171 
\bibitem[\protect\citeauthoryear{Gavazzi et al.}{2002}]%
          {Gavazzi2002} Gavazzi G., Bonfanti C., Sanvito G., Boselli A., Scodeggio M.,  
           2002, ApJ, 576, 135
\bibitem[\protect\citeauthoryear{Goodwin \& Bastian}{2006}]%
          {Goodwin2006} Goodwin S. P., Bastian N., 2006, MNRAS, 373, 752
\bibitem[\protect\citeauthoryear{Gordon et al.}{2008}]%
          {Gordon2008} Gordon K. D., Engelbracht C. W., Rieke G. H., Misselt K. A., 
                       Smith J. -D. T., Kennicutt R. C. Jr., 2008, ApJ, 682, 336
\bibitem[\protect\citeauthoryear{Gilbert et al.}{2000}]%
          {Gilbert2000} Gilbert A. M., Graham J. R., McLean I. S. et al., 2000, ApJ, 533, 57
\bibitem[\protect\citeauthoryear{Gilbert \& Graham}{2007}]%
          {Gilbert2007} Gilbert A. M., Graham J. R., 2007, ApJ, 668, 168
\bibitem[\protect\citeauthoryear{Gil de Paz et al.}{2007}]%
          {GildePaz2007} Gil de Paz A., Boissier S., Madore B. F. et al., 2007, ApJS, 173, 185
\bibitem[\protect\citeauthoryear{Haas, Chini \& Klaas}{2005}]%
          {Haas2005} Haas M., Chini R., Klaas U., 2005, A\&A, 17, L20
\bibitem[\protect\citeauthoryear{Hibbard et al.}{2001}]%
          {Hibbard2001} Hibbard J. E., van der Hulst J. M., Barnes J. E., Rich R. M., 2001, AJ, 122, 2969
\bibitem[\protect\citeauthoryear{Hibbard et al.}{2005}]%
          {Hibbard2005} Hibbard J. E., Bianchi L., Thilker D. A. et al., 2005, ApJ, 619, L87
\bibitem[\protect\citeauthoryear{Holtzman et al.}{1995}]%
          {holtzman1995} Holtzman J. et al., 1995, PASP, 107, 1065
\bibitem[\protect\citeauthoryear{Hummel \& van der Hulst}{1986}]%
          {Hummel1986} Hummel E., van der Hulst J. H., 1986, A\&A, 155, 151
\bibitem[\protect\citeauthoryear{Hughes et al.}{2006}]%
          {hughes2006} Hughes A., Wong T., Ekers R., Staveley-Smith L., Filipovic M., 
                       Maddison S., Fukui Y., Mizuno N., 2006, MNRAS, 370, 363
\bibitem[\protect\citeauthoryear{Jog \& Solomon}{1992}]%
          {Jog1992} Jog C. J., Solomon P. M., 1992, ApJ, 387, 152
\bibitem[\protect\citeauthoryear{Kartaltepe et al.}{2007}]%
          {kartaltepe2007} Kartaltepe J. S., Sanders D. B., Scoville N. Z. et al., 2007, ApJS, 172, 320
\bibitem[\protect\citeauthoryear{Knapen}{1998}]%
          {Knapen1998} Knapen J. H., 1998, MNRAS, 297, 255
\bibitem[\protect\citeauthoryear{Kennicutt et al.}{2007}]%
          {kennicutt2007} Kennicutt R. C., Jr., Calzetti D., Walter F. et al., 2007, ApJ, 671, 333
\bibitem[\protect\citeauthoryear{Kong et al.}{2000}]%
          {kong2000} Kong X., Zhou X., Chen J. S. et al., 2000, AJ, 119, 2745
\bibitem[\protect\citeauthoryear{Kong et al.}{2004}]%
          {kong2004} Kong X., Charlot S., Brinchmann J., Fall S. M., 2004, MNRAS, 349, 769
\bibitem[\protect\citeauthoryear{Kinney et al.}{1993}]%
          {Kinney1993} Kinney A. L., Bohlin R. C., Calzetti D., Panagia N., 
          Wyse R. F. G., 1993, ApJS, 86, 5
\bibitem[\protect\citeauthoryear{Kroupa}{2002}]%
          {Kroupa2002} Kroupa P., 2002, in ASPConf. Ser.285, Modes of Star Formation
           and the Origin of Field Populations, ed.E.K.Grebel \& W.Brandner
           (San Francisco: ASP), 86
\bibitem[\protect\citeauthoryear{Kunze et al.}{1996}]%
          {Kunze1996} Kunze D., Rigopoulou D., Lutz D. et al., 1996, A\&A, 315, L101
\bibitem[\protect\citeauthoryear{Le Fevre et al.}{2000}]%
          {lefevre2000} Le Fevre O., Abraham R., Lilly S. J. et al., 2000, MNRAS, 311, 565
\bibitem[\protect\citeauthoryear{Lin et al.}{2008}]%
           {lin2008} Lin L., Patton D. R., Koo D. C. et al., 2008, ApJ, 681, 232
\bibitem[\protect\citeauthoryear{Lu et al.}{1996}]%
          {lu1996} Lu N. Y., Helou G., Tuffs R., Xu C., Malhotra S., 
                   Werner M. W., Thronson H., 1996, A\&A, 315, L153
\bibitem[\protect\citeauthoryear{Martin-Hernandez et al.}{2000}]%
          {martin2000} Martin-Hernandez N. L., Peeters E., Damour F. et al., 2000, ESASP, 456, 135
\bibitem[\protect\citeauthoryear{Mirabel et al.}{1998}]%
          {Mirabel1998} Mirabel L. F., Vigroux L., Charmandaris V. et al., 1998, A\&A, 333, L1
\bibitem[\protect\citeauthoryear{Mengel et al.}{2001}]%
          {Mengel2001} Mengel S., Lehnert M. D., Thatte N., Tacconi-Garman L. E.,
           Genzel R. 2001, ApJ, 550, 280
\bibitem[\protect\citeauthoryear{Mengel et al.}{2002}]%
          {Mengel2002} Mengel S., Lehnert M. D., Thatte N., Tacconi-Garman L. E.,
          Genzel R., 2002, ApJ, 383, 137
\bibitem[\protect\citeauthoryear{Mengel et al.}{2005}]%
          {Mengel2005} Mengel S., Lehnert M. D., Thatte N., Genzel, R., 
          2005, A\&A, 443, 41
\bibitem[\protect\citeauthoryear{Mihos, Bothun \& Richstone}{1993}]%
          {Mihos1993} Mihos J. C., Bothun G. D., Richstone D. O., 1993, ApJ, 418, 82
\bibitem[\protect\citeauthoryear{Mihos \& Hernquist}{1996}]%
          {Mihos1996} Mihos J. C., Hernquist L., 1996, ApJ, 464, 641
\bibitem[\protect\citeauthoryear{Murphy et al.}{2006}]%
          {murphy2006} Murphy E. J., Braun R., Helou G. et al., 2006, ApJ, 638, 157 
\bibitem[\protect\citeauthoryear{Nikola et al.}{1998}]%
          {Nikola1998} Nikola T., Genzel R., Herrmann F., Madden S. C., Poglitsch A., 
                       Geis N., Townes C. H., Stacey G. J., 1998, ApJ, 504, 749
\bibitem[\protect\citeauthoryear{Neff \& Ulvestad}{2000}]%
          {Neff2000} Neff S. G., Ulvestad J. S., 2000, AJ, 120, 670
\bibitem[\protect\citeauthoryear{Oey et al.}{2003}]%
          {Oey2003} Oey M. S., Parker J. S., Mikles V. J., Zhang X. L., 2003, AJ, 126, 2317
\bibitem[\protect\citeauthoryear{Peeters et al.}{2002}]%
          {peeters2002} Peeters E., Martin-Hernandez N. L., Damour F. et al., 2002, A\&A, 381, 571
\bibitem[\protect\citeauthoryear{Peeters et al.}{2002}]%
          {patton2002} Patton D. R., Pritchet C. J., Carlberg R. G. et al., 2002, ApJ, 565, 208
\bibitem[\protect\citeauthoryear{Read, Ponman \& Wolstencraft}{1995}]%
          {Read1995} Read A. M., Ponman T. J., Wolstencraft R. D., 1995, MNRAS, 277, 397
\bibitem[\protect\citeauthoryear{Rieke et al.}{2004}]%
          {Rieke2004} Rieke G. H., Young E. T., Engelbracht C. W. et al., 2004, ApJS, 154, 25
\bibitem[\protect\citeauthoryear{Sanders \& Mirabel}{1996}]%
          {Sanders1996} Sanders D. B., Mirabel I. F., 1996, ARA\&A, 34, 749
\bibitem[\protect\citeauthoryear{Sanders et al.}{2003}]%
          {Sanders2003} Sanders D. B., Mazzarella J. M., Kim D. -C., Surace J. A., 
                        Soifer B. T., 2003, ApJ, 126, 1607
\bibitem[\protect\citeauthoryear{Saviane et al.}{2008}]%
           {saviane2008} Saviane I., Momany Y., da Costa G. S., Rich R. M., Hibbard J. E. 2008, 
                        ApJ, 678, 179
\bibitem[\protect\citeauthoryear{Schweizer et al.}{2008}]%
            {schweizer2008} Schweizer F., Burns C. R., Madore B. F., et al., 2008, AJ, 136, 1482 
\bibitem[\protect\citeauthoryear{Schlegel, Finkbeiner \& Davis}{1998}]%
          {Schlegel1998} Schlegel D. J., Finkbeiner D. P., Davis M., 1998, ApJ, 500, 525
\bibitem[\protect\citeauthoryear{Schulz et al.}{2007}]%
          {Schulz2007} Schulz A., Henkel C., Muders D., Mao R. Q., Rollig M., 
                       Mauersberger R., 2007, A\&A, 466, 467
\bibitem[\protect\citeauthoryear{Smith et al.}{2007}]%
          {Smith2006} Smith J. D. T., Draine B. T., Dale D. A. et al., 2007, ApJ, 656, 770
\bibitem[\protect\citeauthoryear{Stetson}{1987}]%
          {Stetson1987} Stetson P. B. 1987, PASP, 99, 191
\bibitem[\protect\citeauthoryear{Thilker et al.}{2007}]%
          {Thilker2007b} Thilker D. A., Boissier S., Bianchi L. et al., 2007, ApJ, 173, 572 
\bibitem[\protect\citeauthoryear{Toomre \& Toomre}{1972}]%
          {Toomre1972} Toomre A., Toomre J., 1972, ApJ, 178, 623
\bibitem[\protect\citeauthoryear{Vazquez \& Leitherer}{2005}]%
          {Vazquez2005} Vazquez G. A., Leitherer C., 2005, ApJ, 621, 695
\bibitem[\protect\citeauthoryear{Vigroux et al.}{1996}]%
          {Vigroux1996} Vigroux L., Mirabel F., Altieri B. et al., 1996, A\&A, 315, L93
\bibitem[\protect\citeauthoryear{Wang et al.}{2004}]%
          {Wang2004} Wang Z., Fazio G. G., Ashby M. L. N. et al., 2004, ApJS, 154, 193
\bibitem[\protect\citeauthoryear{Wilson et al.}{2000}]%
          {Wilson2000} Wilson C. D., Scoville N., Madden S. C., Charmandaris V. 2000, 
          ApJ, 542, 120
\bibitem[\protect\citeauthoryear{Wilson et al.}{2003}]%
          {Wilson2003} Wilson C. D., Scoville N., Madden S. C., Charmandaris V. 2003, 
          ApJ, 599, 1049
\bibitem[\protect\citeauthoryear{Whitmore \& Schweizer}{1995}]%
          {whitmore1995} Whitmore B., Schweizer F., 1995, AJ, 109, 960
\bibitem[\protect\citeauthoryear{Whitmore et al.}{1999}]%
         {Whitmore1999}Whitmore B. C, Zhang Q., Leitherer C., Fall S. M, Schweizer F., 
          Miller B. W., 1999, AJ, 118, 1551
\bibitem[\protect\citeauthoryear{Whitmore \& Zhang}{2002}]%
          {Whitmore2002} Whitmore B. C, Zhang Q., 2002, AJ, 124, 1418
\bibitem[\protect\citeauthoryear{Whitmore}{2004}] %
          {Whitmore2004} Whitmore B. C., 2004, in The Formation and Evolution of Massive 
          Young Star Clusters, ASPC, 322, 419
\bibitem[\protect\citeauthoryear{Yun et al.}{2001}] %
          {yun2001} Yun M. S., Reddy N. A., Condon J. J., 2001, ApJ, 554, 803
\bibitem[\protect\citeauthoryear{Zhang et al.}{2001}] %
          {zhang2001} Zhang Q., Fall M., Whitmore B. C., 2001, ApJ, 561, 727
\bibitem[\protect\citeauthoryear{Zhu et al.}{2007}] %
          {zhu2007} Zhu M., Gao Y., Seaquist E. R., 2007, AJ, 134, 118
\end{thebibliography}
\end{document}